\documentclass[aps,prd,twocolumn,nofootinbib]{revtex4-1}

\usepackage{graphicx}

\begin{document}

\title{New results on Pionic Twist-3 Distribution Amplitudes within the QCD Sum Rules}
\author{Tao Zhong}
\author{Xing-Gang Wu}
\email{wuxg@cqu.edu.cn}
\author{Jia-Wei Zhang}
\author{Yun-Qing Tang}
\author{Zhen-Yun Fang}
\affiliation{Department of Physics, Chongqing University, Chongqing
400044, P.R. China}

\date{\today}

\begin{abstract}
We present an improved calculation on the pionic twist-3 distribution amplitudes $\phi^{\pi}_{p}$ and $\phi^{\pi}_{\sigma}$, which are studied within the QCD sum rules. By adding all the uncertainties in quadrature, it is found that $\left<\xi^2_p\right>=0.248^{+0.076}_{-0.052}$, $\left<\xi^4_p\right>=0.262^{+0.080}_{-0.055}$, $\left<\xi^2_\sigma\right>=0.102^{+0.035}_{-0.025}$ and $\left<\xi^4_\sigma\right>=0.094^{+0.028}_{-0.020}$. Furthermore, with the help of these moments, we construct a model for the twist-3 wave functions $\psi^{\pi}_{p,\sigma}(x,\mathbf{k}_\bot)$, which have better end-point behavior and are helpful for perturbative QCD approach. The obtained twist-3 distribution amplitudes are adopted to calculate the $B\to\pi$ transition form factor $f^+_{B\pi}$ within the QCD light-cone sum rules up to next-to-leading order. By suitable choice of the parameters, we obtain a consistent $f^+_{B\pi}$ with those obtained in the literature. \\



\end{abstract}

\maketitle

\section{Introduction}

Distribution amplitude (DA) shows the momentum fraction distributions of partons in hadron in a particular Fock state, which posses an important component for QCD factorization theory \cite{brodsky}. It is convenient to arrange DA by its different twist structures such that when the energy scale of the process is large enough, one can safely neglect those higher power suppressed contributions from higher twists. Due to its simpler structure, the leading-twist DA has attracted much attention in the literature \cite{brodsky,cz}. Very recently, the new BABAR data on the pion-photon transition form factor \cite{babar} arouses people's new interests on the pion twist-2 DA \cite{twist2DA}.

However the higher-twist DAs are much more involved \cite{pballt,NPB529,JHEP9901}, which describes either contributions of the transverse motion of quarks (antiquarks) in the leading-twist components or contributions of higher Fock states with additional gluons and/or quark-antiquark pairs or etc.. In certain cases, higher twist structures especially the twist-3 DAs may provide sizable contributions and there are usually the main uncertainties for the theoretical estimation. For example, as for the case of pion electro-magnetic form factor and $B\to\pi$ transition form factor, most calculations give large twist-3 contributions that are even dominant over that of the leading twist in the intermediate or even large $Q^2$ region. It should be pointed out that those large contributions of twist-3 DAs are usually based on the asymptotic behavior of $\phi^{\pi}_p$, i.e. $\phi^{AS}_p\equiv1$. And with such a naive asymptotic behavior of twist-3 DA, the end-point singularity can not be effectively suppressed, which inversely leads to a large twist-3 contribution \cite{hw1,hw2,hw3,hw4,hw5}. While, by taking the $k_T$ factorization approach \cite{kt} and by constructing a twist-3 wavefunction model based on the DA moments derived by Refs.\cite{EPJC42,PRD70}, Refs.\cite{piff,bpiff} show that even though in the intermediate energy region, both leading twist and twist-3 DAs can provide sizable contributions, however the contributions from twist-3 ones are really power suppressed for large $Q^2$ region, and then the normal power counting rule is retrieved. So a twist-3 DA with a better end-point behavior other than the asymptotic one shall lead to a better understanding of these form factors.

At the present, the pionic twist-3 structures are far from affirmation \cite{YF41,ZPC48,NPB529,JHEP9901,EPJC42,PRD70}, and it would be interesting to do further studies on the twsit-3 DAs / wavefunctions. In the present paper, we shall first make a study on the pionic twist-3 DA moments within the QCD sum rules, and then make a discussion on its physical effects by constructing a reasonable wavefunction model and by taking $B\to\pi$ transition form factor as an explicit example.

The remaining parts of the paper is organized as follows. In Sec.II, we present the calculation technology for the twist-3 DAs $\phi^{\pi}_{p}$ and $\phi^{\pi}_{\sigma}$ within the framework of QCD sum rules. In Sec.III, we present our numerical results for $\phi^{\pi}_{p}$ and $\phi^{\pi}_{\sigma}$, where the DA moments together with their uncertainties are discussed in detail. Based on the derived DA moments, we construct the wavefunction models for $\psi^{\pi}_{p}$ and $\psi^{\pi}_{\sigma}$ in Sec.IV, and then make a discussion on their rationality by further applying them to deal with the $B\to\pi$ transition form factor within the QCD light-cone sum rules up to next-to-leading order (NLO). The final section is reserved for a summary.

\section{Calculation Technology for the Twist-3 DAs $\phi^{\pi}_{p}$ and $\phi^{\pi}_{\sigma}$}

The pionic twist-3 DAs $\phi^{\pi}_{p}$ and $\phi^{\pi}_{\sigma}$ are defined as \cite{ZPC48}
\begin{eqnarray}
\left< 0 \left| \bar{u}(x) i\gamma_5 d(-x) \right| \pi^-_q \right>=\mu_{\pi} f_{\pi} \int^1_0 du e^{i\xi (q \cdot x)} \phi^\pi_p(u)
\end{eqnarray}
and
\begin{widetext}
\begin{equation}
\left< 0 \left| \bar{u}(x) \sigma_{\mu\nu} \gamma_5 d(-x) \right| \pi^-_q \right> = -\frac{i}{3} \mu_{\pi}f_{\pi} \int^1_0 du e^{i\xi (q \cdot x)}(q_\mu x_\nu - q_\nu x_\mu) \phi^\pi_\sigma(u),
\end{equation}
\end{widetext}
where $\xi = (2u-1)$, and due to the equations of motion of the quarks inside pion \cite{ZPC48}, $\mu_{\pi}=\frac{m^2_\pi}{m_u + m_d}$, whose value is $1.6\pm 0.2\rm GeV$ at the scale $1$ GeV \cite{JHEP9901}. Normally, these two DAs can be expanded by the conventional Gegenbauler expansions, $\phi^{\pi}_{p}(\xi)=\sum_{n}\left< \xi^{2n}_p \right> C^{1/2}_{2n}(\xi)$ and $\phi^{\pi}_{\sigma}(\xi)=3(1-\xi^2)/4\sum_{n} \left< \xi^{2n}_\sigma\right> C^{3/2}_{2n}(\xi)$, where $C^{1/2,3/2}_{2n}(\xi)$ are Gegenbauler polynomials and $\left< \xi^{2n}_{p,\sigma} \right>$ are the so-called Gegenbauler moments. As for the pionic case, because of the chiral symmetry, only even moments' terms are non-zero.

The Gegenbauler moments can be calculated under the QCD sum rules. For such purpose, one can define the two correlation functions as
\begin{widetext}
\begin{equation}
(z \cdot q)^{2n} I^{(2n,0)}_{p} (q^{2}) \equiv -i \int d^{4} x e^{iq \cdot
x} \left< 0 \left| T \{ \bar{d}( x ) \gamma_{5} ( i z \cdot
\tensor{D})^{2n} u(x) , \bar{u}( 0 ) \gamma_{5} d(0) \} \right| 0 \right>
\end{equation}
and
\begin{equation}
- i (q_\mu z_\nu - q_\nu z_\mu) (z \cdot q)^{2n} I^{(2n,0)}_\sigma (q^2) \equiv -i \int d^4 x e^{i q \cdot x} \left< 0 \left| T \{ \bar{d}(x) \sigma_{\mu\nu} \gamma_5 ( i z \cdot \tensor{D} )^{2n+1} u(x) , \bar{u}(0) \gamma_5 d(0) \} \right| 0 \right>.
\end{equation}
\end{widetext}
Following the same calculation technology as described in detail in Refs.\cite{EPJC42,PRD70}, i.e. the QCD sum rules under the background field approach \cite{bkg1,bkg2,bkg3,bkg4}, we obtain two sum rules for the moments of $\phi^{\pi}_p$ and $\phi^{\pi}_\sigma$ with the condensates up to dimension six, i.e.
\begin{widetext}
\begin{eqnarray}
\left< \xi^{2n}_p \right> &=& \frac{M^4 e^{m^2_{\pi} / M^2}}{f^2_\pi \mu_\pi^2}\left\{ \frac{3}{8 \pi^2} \frac{1}{2n+1} \left[ 1 - \left( 1 + \frac{s^p_\pi}{ M^2} \right) e^{-s^{p}_{\pi} / M^{2}} \right] + \frac{2n-1}{2}\frac{(m_u + m_d) \left< \bar{q} q \right> }{M^4} \right. \nonumber\\
&&\left. + \frac{2n+3}{24} \frac{ \left< \frac{\alpha_s}{\pi} G^2 \right>}{M^4} + \frac{16\pi}{81} [21+8n(n+1)] \frac{\left< \sqrt{\alpha_s} \bar{q} q \right>^2}{M^6} \right\} \label{srp}
\end{eqnarray}
and
\begin{eqnarray}
\left< \xi^{2n}_\sigma \right> &=& \frac{M^4 e^{m^2_{\pi} / M^2}}{f^2_\pi \mu_\pi^2} \frac{3}{2n+1} \left\{ \frac{3}{8 \pi^2} \frac{1}{2n+3} \left[ 1 - \left( 1 + \frac{s^{\sigma}_\pi}{M^2} \right) e^{-s^{\sigma}_{\pi} / M^{2}} \right]+ \frac{2n+1}{2}\frac{(m_u + m_d) \left< \bar{q} q \right>}{M^4} \right. \nonumber\\
&& \left.  + \frac{2n+1}{24} \frac{ \left< \frac{\alpha_s}{\pi} G^2 \right>}{M^4} + \frac{16\pi}{81} (8n^2 - 2) \frac{\left< \sqrt{\alpha_s} \bar{q} q \right>^2}{M^6} \right\}, \label{srsigma}
\end{eqnarray}
\end{widetext}
where $M$ is the Borel parameter, $s^{p}_{\pi}$ and $s^{\sigma}_{\pi}$ are continuum threshold, which are usually taken to be around the mass square of the first exciting state $\pi'(1300)$ of pion. For clarity, we will take $ s_\pi^{p,\sigma}=1.59$, $1.69$ and $1.79$ ${\rm GeV}^2$ to do our discussion. For the non-perturbative vacuum condensate, we take \cite{condensates}: $\left< \bar{u} u \right> = \left< \bar{d} d \right>  \simeq -(240{\rm ~MeV})^3$ and $\left< {\displaystyle\frac{\alpha_s}{\pi}} G^2 \right> = 0.012{\rm ~GeV}^4$, which are at the renormalization scale $1$ GeV and shall be evaluated up to the concerned scale by using the leading order anomalous dimensions of the condensates. The pion decay constant, the pion mass and the $\alpha_s$ are taken as the center values of Refs.\cite{pdg,alphas}, i.e., $f_\pi = 0.1304{\rm ~GeV}$ and $m_\pi = 0.1396{\rm ~GeV}$ and $\alpha_s(M_Z)=0.1184$. The renormalization scale for the present case is taken as $\mu = M$.

\begin{figure}
\includegraphics[width=0.45\textwidth]{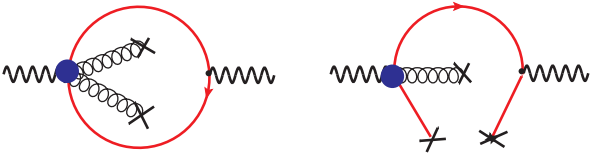}
\caption{Feynman diagrams that are missed in Ref.\cite{EPJC42}, where the gluon and the quark condensates are depicted as crosses, and the big dot stands for the currents $\left[\bar{d}(x)\gamma_{5}(i z\cdot \tensor{D}
)^{2n}u(x)\right]$ and $\left[\bar{d}(x)\sigma_{\mu\nu}\gamma_5 (i z\cdot \tensor{D} )^{2n+1}u(x)\right]$ for $\phi^\pi_p$ and $\phi^{\pi}_{\sigma}$ respectively. }
\label{fig1}
\end{figure}

As a cross check, we make a comparison with the sum rules derived in Refs.\cite{EPJC42,PRD70}. Firstly, our sum rules agree with the corresponding sum rules derived by Ref.\cite{EPJC42}, except for the two coefficients of the dimension-four $\left< \frac{\alpha_s}{\pi} G^2 \right>$ and dimension-six matrix element $\left< \sqrt{\alpha_s} \bar{q} q \right>^2$. It is found that such a difference is caused by the two Feynman diagrams shown by FIG.(\ref{fig1}), which are missed by the authors of Ref.\cite{EPJC42}. According to the background field approach \cite{bkg1,bkg2,bkg3,bkg4}, the Feynman rules are derived under the Schwinger Gauge or `the fixed-point gauge' \cite{sch}. And because of the gluon's or quark's equation of motion, only those interactions that have at most one background field coupling can have contribution to the interaction vertex. As for the two Feynman diagrams shown by FIG.(\ref{fig1}), the two background fields are attached to a big dot, which stands for the two complex currents $\left[\bar{d}(x)\gamma_{5}(i z\cdot \tensor{D})^{2n}u(x)\right]$ and $\left[\bar{d}(x)\sigma_{\mu\nu}\gamma_5 (i z\cdot \tensor{D})^{2n+1}u(x)\right]$ for $\phi^\pi_p$ and $\phi^{\pi}_{\sigma}$ respectively. Since the two background fields can attach to different places of the two complex currents, these two Feynman should be taken into consideration. Secondly, Ref.\cite{PRD70} only presents the sum rules for $\phi_p$. And, one may find a typo error in Ref.\cite{PRD70}, i.e. the term before the dimension-three quark condensate $<\bar q q>$ and dimension-six quark condensate $< \bar{q} q>^2$ should be $``+"$ other than $``-"$.

\section{Numerical Results for the Moments of $\phi^{\pi}_p$ and $\phi^{\pi}_\sigma$}

Basing on the QCD sum rules (\ref{srp}) and (\ref{srsigma}), we discuss the Gegenbauler moments of  $\phi^{\pi}_{p}$ and $\phi^{\pi}_{\sigma}$. For the purpose, we adopt the usual criteria to fix the Borel window for the first two moments $\left<\xi^{2,4}_{p,\sigma}\right>$, i.e. the continuum contribution is less than $30\%$ of the total dispersion integration and the dimension-six condensate contribution is no more than $30\%$. It is found that the continue contribution increases with the {\it increment} of $M^2$ and the contribution from the dimension-six condensate term increases with the {\it decrement} of $M^2$, then possible Borel windows can be obtained as required.

\begin{table}
\caption{Uncertainties of the second moment $\left<\xi^2_p\right>_{\mu_\pi}$, where $\mu_\pi (1{\rm GeV}) = 1.4$, $1.6$, $1.8$ GeV respectively. }
\begin{tabular}{|c|c|c|c|}
\hline \hline
$s^p_\pi(\rm GeV^2)$ & $1.59$ & $1.69$ & $1.79$\\
\hline \hline
$M^2(\rm GeV^2)$ & $[0.696,0.874]$ & $[0.696,0.905]$ & $[0.696,0.937]$\\
$\left<\xi^2_p\right>_{(1.4)}$ & $0.315 \pm 0.002$ & $0.324 \pm 0.005$ & $0.334 \pm 0.009$\\
$\left<\xi^2_p\right>_{(1.6)}$ & $0.242 \pm 0.002$ & $0.248 \pm 0.004$ & $0.255 \pm 0.007$\\
$\left<\xi^2_p\right>_{(1.8)}$ & $0.191 \pm 0.001$ & $0.196 \pm 0.003$ & $0.202 \pm 0.005$\\
\hline \hline
\end{tabular}
\label{tabp1}
\end{table}

\begin{table}
\caption{Uncertainties of the fourth moment $\left<\xi^{4}_p\right>_{\mu_\pi}$, where $\mu_\pi (1{\rm GeV}) = 1.4$, $1.6$, $1.8$ GeV respectively. }
\begin{tabular}{|c|c|c|c|}
\hline \hline
$s^p_\pi(\rm GeV^2)$ & $1.59$ & $1.69$ & $1.79$\\
\hline \hline
$M^2(\rm GeV^2)$ & $[0.958,1.052]$ & $[0.958,1.080]$ & $[0.958,1.108]$\\
$\left<\xi^4_p\right>_{(1.4)}$ & $0.335 \pm 0.004$ & $0.342 \pm 0.004$ & $0.349 \pm 0.004$\\
$\left<\xi^4_p\right>_{(1.6)}$ & $0.257 \pm 0.003$ & $0.262 \pm 0.003$ & $0.267 \pm 0.003$\\
$\left<\xi^4_p\right>_{(1.8)}$ & $0.203 \pm 0.002$ & $0.207 \pm 0.002$ & $0.211 \pm 0.002$\\
\hline \hline
\end{tabular}
\label{tabp2}
\end{table}

\begin{table}
\caption{Uncertainties of the second moment $\left<\xi^2_\sigma\right>_{\mu_\pi}$, where $\mu_\pi (1{\rm GeV}) = 1.4$, $1.6$, $1.8$ GeV respectively. }
\begin{tabular}{|c|c|c|c|}
\hline \hline
$s^\sigma_\pi(\rm GeV^2)$ & $1.59$ & $1.69$ & $1.79$\\
\hline \hline
$M^2(\rm GeV^2)$ & $[0.439,0.798]$ & $[0.439,0.832]$ & $[0.439,0.867]$\\
$\left<\xi^2_\sigma\right>_{(1.4)}$ & $0.129 \pm 0.013$ & $0.134 \pm 0.017$ & $0.139 \pm 0.021$\\
$\left<\xi^2_\sigma\right>_{(1.6)}$ & $0.099 \pm 0.010$ & $0.102 \pm 0.013$ & $0.106 \pm 0.016$\\
$\left<\xi^2_\sigma\right>_{(1.8)}$ & $0.078 \pm 0.008$ & $0.081 \pm 0.010$ & $0.084 \pm 0.013$\\
\hline \hline
\end{tabular}
\label{tabs1}
\end{table}

\begin{table}
\caption{Uncertainties of the fourth moment $\left<\xi^4_\sigma\right>_{\mu_\pi}$, where $\mu_\pi (1{\rm GeV}) = 1.4$, $1.6$, $1.8$ GeV respectively. }
\begin{tabular}{|c|c|c|c|}
\hline \hline
$s^\sigma_\pi(\rm GeV^2)$ & $1.59$ & $1.69$ & $1.79$\\
\hline \hline
$M^2(\rm GeV^2)$ & $[0.807,0.988]$ & $[0.807,1.017]$ & $[0.807,1.047]$\\
$\left<\xi^4_\sigma\right>_{(1.4)}$ & $0.120 \pm 0.001$ & $0.122 \pm 0.001$ & $0.125 \pm 0.001$\\
$\left<\xi^4_\sigma\right>_{(1.6)}$ & $0.092 \pm 0.001$ & $0.094 \pm 0.001$ & $0.096 \pm 0.001$\\
$\left<\xi^4_\sigma\right>_{(1.8)}$ & $0.072 \pm 0.001$ & $0.074 \pm 0.001$ & $0.076 \pm 0.001$\\
\hline \hline
\end{tabular}
\label{tabs2}
\end{table}

\begin{figure}
\begin{center}
\includegraphics*[width=0.4\textwidth]{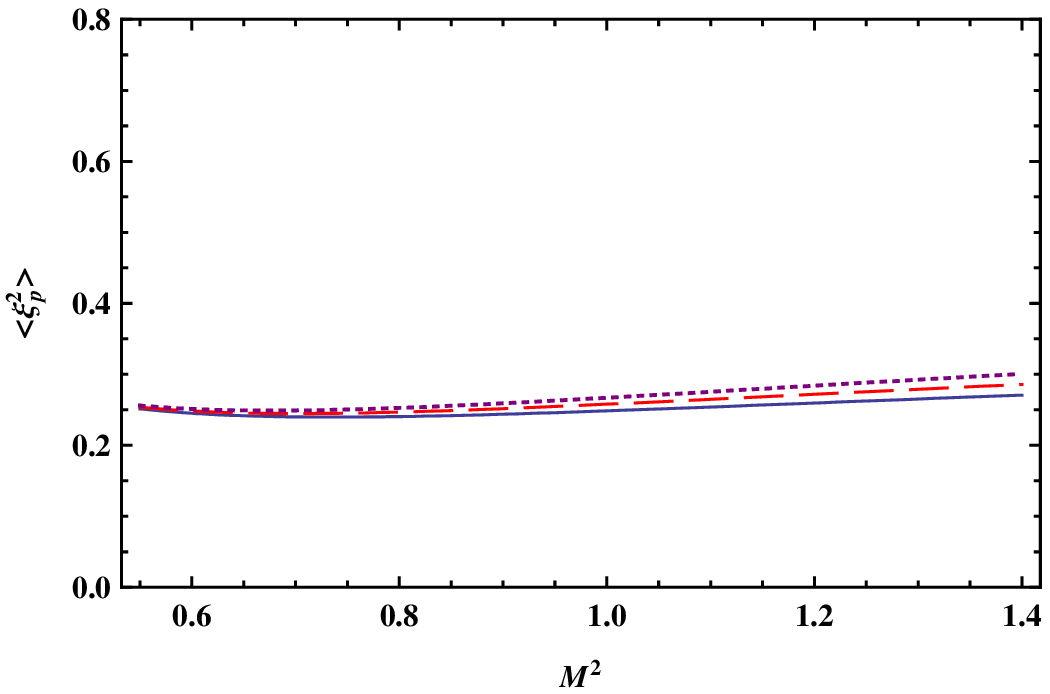} \includegraphics*[width=0.4\textwidth]{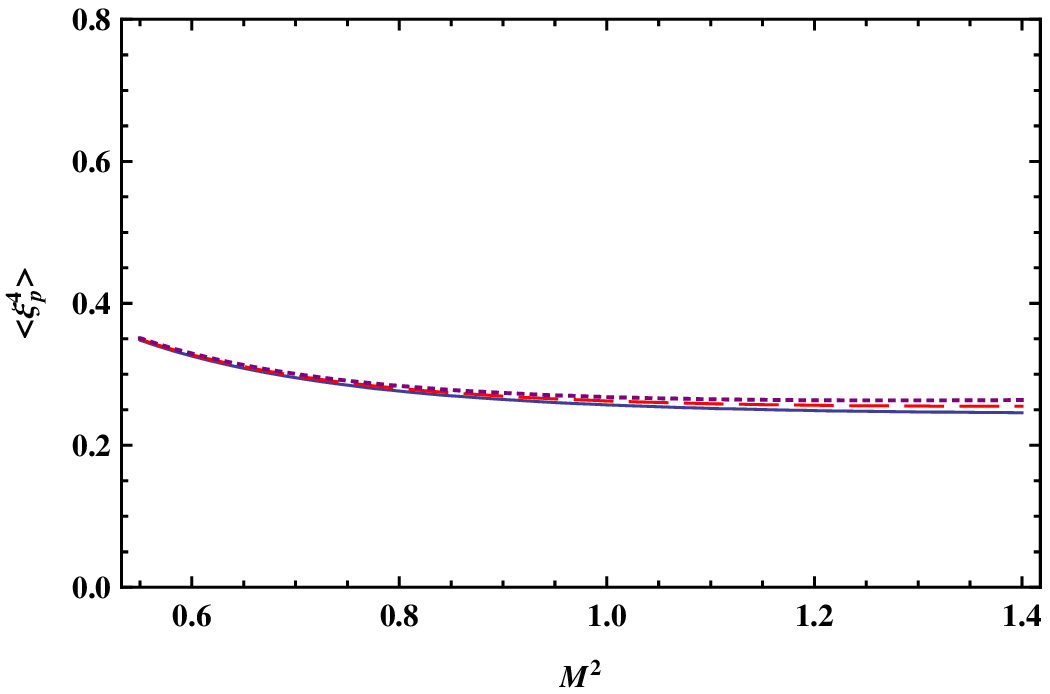}
\end{center}
\caption{The second and forth moments of twist-3 DA $\phi^\pi_{p}$ versus the Borel parameter $M^2$ corresponding to $\mu_\pi = 1.6\rm GeV$. The solid line, dashed line and dotted line correspond $s^{p}_\pi=1.59$, $1.69$, $1.79 {\rm GeV}^2$, respectively.}
\label{momentsp}
\end{figure}

\begin{figure}
\begin{center}
\includegraphics*[width=0.4\textwidth]{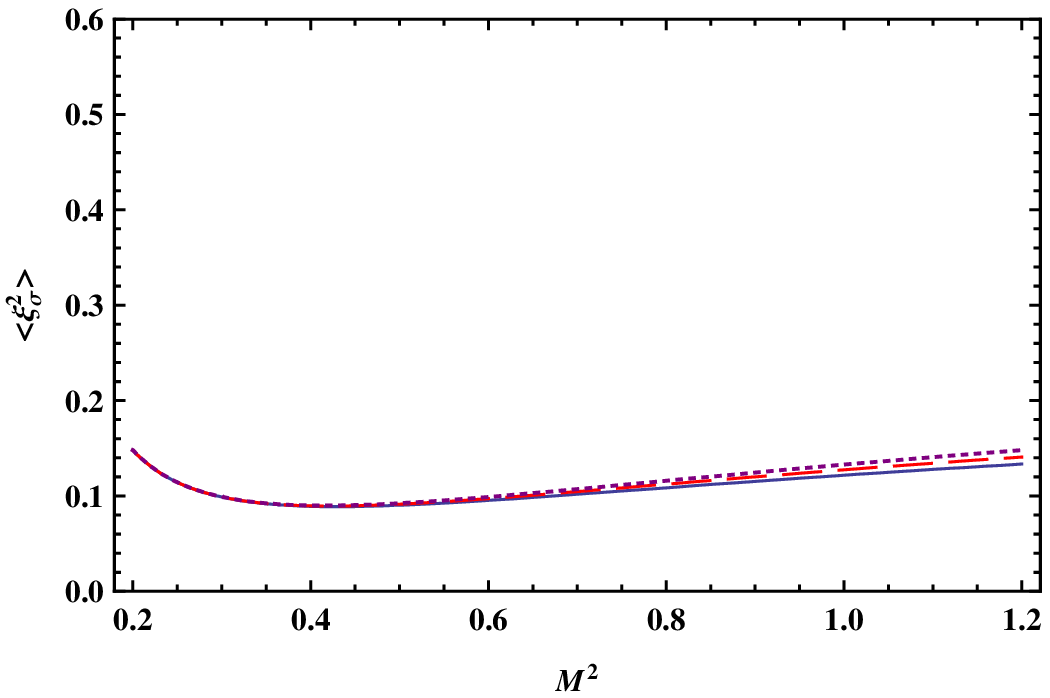} \includegraphics*[width=0.4\textwidth]{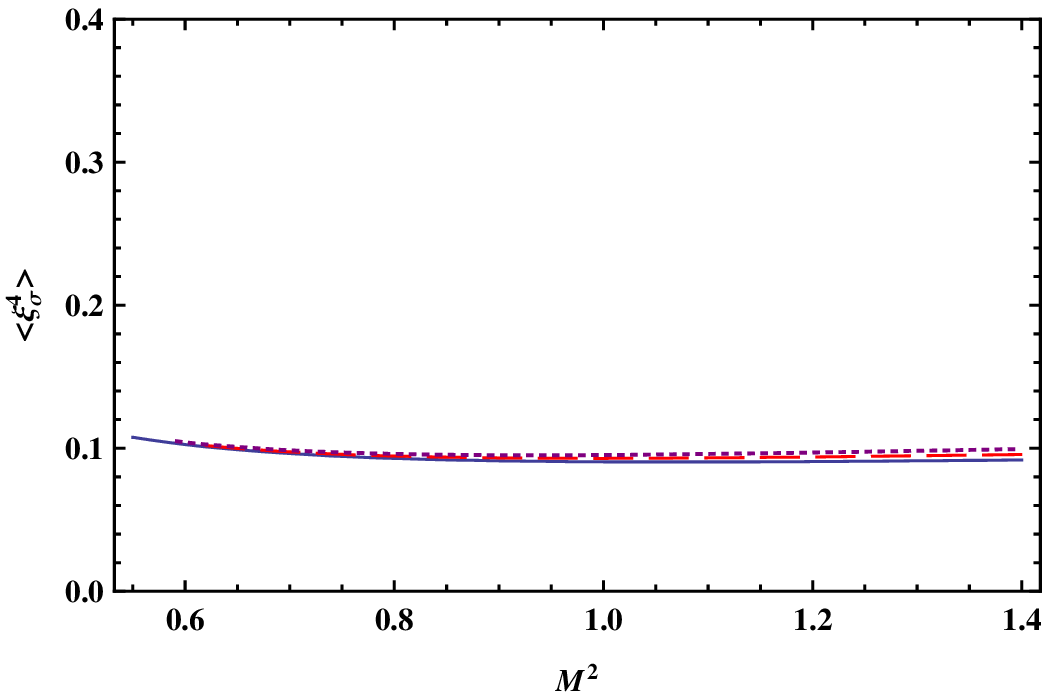}
\end{center}
\caption{The second and forth moments of twist-3 DAs $\phi^\pi_{\sigma}$ versus the Borel parameter $M^2$ corresponding to $\mu_\pi = 1.6\rm GeV$. The solid line, dashed line and dotted line correspond $s^{\sigma}_\pi=1.59$, $1.69$, $1.79 {\rm GeV}^2$, respectively.}
\label{momentss}
\end{figure}

All the obtained Borel windows and the corresponding second and fourth moments of $\phi^{\pi}_{p}$ and $\phi^{\pi}_{\sigma}$ are collected in TABs.(\ref{tabp1},\ref{tabp2},\ref{tabs1},\ref{tabs2}) respectively. TABs.(\ref{tabp1},\ref{tabp2},\ref{tabs1},\ref{tabs2}) indicate that

\begin{itemize}
\item The second and fourth moments of $\phi^{\pi}_p$ are slightly affected by the Borel parameter $M^2$. This shows that the present sum rules (\ref{srp}) is reasonable. More explicitly, it can be found that the second and forth moments of $\phi^{\pi}_{p}$, the uncertainty caused by $M^2$ is less than $2\%$. For sum rules (\ref{srsigma}), it is noted that for the second moment of $\phi^{\pi}_{\sigma}$, the uncertain caused by $M^2$ reaches up to a somewhat larger value $\sim13\%$. While for higher moments, it is found that the uncertainty caused by $M^2$ is back to be less than $2\%$. If we require the flatness of the moments versus $M^2$ as an extra condition to constrain the Borel window then the large uncertainty for the second moments can be reduced. Here to make a consistent analysis over all the moments of $\phi^{\pi}_{p}$ and $\phi^{\pi}_{\sigma}$, we do not imply any further constraint for $\phi^{\pi}_{\sigma}$. To show these points more clearly, we draw the curves of the second and forth moments of $\phi^\pi_{p,\sigma}$ versus $M^2$ in FIGs.(\ref{momentsp},\ref{momentss}), which shows that the moments (except for the second moments of $\phi^{\pi}_{\sigma}$) are almost flat in the allowable Borel window.

\item The Borel windows will be broadened with the increment of the continue threshold $s^{p,\sigma}_\pi$. And accordingly, the second and the fourth moments of $\phi^{\pi}_{p,\sigma}$ shall increase with increment of $s^{p,\sigma}_\pi$. When increasing $s^{p,\sigma}_\pi$ by $0.10$ GeV, the second and fourth moments of $\phi^{\pi}_{p,\sigma}$ shall be increased by $3\%-5\%$.

\item The twist-3 DA moments are greatly affected by the value of $\mu_\pi$, which provides the dominant uncertainty for the present sum rules. So a better determination of $\mu_\pi$ shall greatly improve our knowledge on twist-3 DA. All the twist-3 DA moments are decreased with the increment of $\mu_\pi$. When increasing $\mu_\pi$ by $0.2$ GeV, the second and fourth moments of $\phi^{\pi}_{p,\sigma}$ shall be decreased by $20\%-25\%$.

\item By taking all uncertainty sources into consideration, we obtain
    \begin{eqnarray}
    \left<\xi^2_p\right>&=&0.248\pm0.004|_{M^2} {\ }^{+0.007}_{-0.006}|_{s^p_\pi} {\ }_{-0.052}^{+0.076}|_{\mu_{\pi}} , \nonumber\\
    \left<\xi^4_p\right>&=&0.262\pm0.003|_{M^2} {\ }^{+0.005}_{-0.005}|_{s^p_\pi} {\ }_{-0.055}^{+0.080}|_{\mu_{\pi}} , \nonumber\\
    \left<\xi^2_\sigma\right>&=&0.102\pm0.013|_{M^2} {\ }^{+0.004}_{-0.003}|_{s^\sigma_\pi} {\ }_{-0.021}^{+0.032}|_{\mu_{\pi}}
    \end{eqnarray}
    and
    \begin{equation}
    \left<\xi^4_\sigma\right> = 0.094\pm0.001|_{M^2} {\ }^{+0.002}_{-0.002}|_{s^\sigma_\pi} {\ }_{-0.020}^{+0.028}|_{\mu_{\pi}} ,
    \end{equation}
    where the center value are obtained by taking all the input parameters to be their center values, i.e. $s^{p,\sigma}_\pi=1.69$ GeV$^2$ and $\mu_\pi(1GeV)=1.6$ GeV. And the uncertain of a particular parameter is obtained by fixing other parameters to be their center accordingly.

\item As a summary, by adding the uncertainties in quadrature, we obtain $\left<\xi^2_p\right>=0.248^{+0.076}_{-0.052}$, $\left<\xi^4_p\right>=0.262^{+0.080}_{-0.055}$, $\left<\xi^2_\sigma\right>=0.102^{+0.035}_{-0.025}$ and $\left<\xi^4_\sigma\right>=0.094^{+0.028}_{-0.020}$.
\end{itemize}

\section{Models for $\psi^{\pi}_{p}$ and $\psi^{\pi}_{\sigma}$ and Their Application for $B\to\pi$ Form Factor}

\subsection{Models for $\psi^{\pi}_{p}$ and $\psi^{\pi}_{\sigma}$}

The wavefunction is the key component for the $k_T$ factorization approach \cite{kt}, which is helpful to suppress the end-point singularity in certain processes. And it would be useful to know the properties of the twist-3 wavefunctions $\psi^{\pi}_{p}$ and $\psi^{\pi}_{\sigma}$.

The wavefunction is purely nonperturbative, so it can hardly be determined from the first principle of QCD. Generally, the pion wavefunction and its corresponding DA is related through the relation, $\phi^\pi_{p,\sigma}(x,\mu_f)=\int_{|k_\bot|< \mu_{f}} \frac{d^2\mathbf{k}_\bot}{16\pi^3}\psi^\pi_{p,\sigma}(x,\mathbf{k}_\bot)$, where $\mu_f$ is the factorization scale that is around ${\cal O}(1 {\rm GeV})$. So, on the other hand, if we have known the DA moments well, then we can inversely obtain some valuable properties of the wavefunction.

Our model for the pionic twsit-3 wavefunctions is based on the assumptions that I) its longitudinal distributions is determined by its DA moments as derived by the sum rules (\ref{srp}) and (\ref{srsigma}). And its longitudinal behavior is dominated by the first two Gegenbauler moments; II) its transverse momentum dependence is determined by the so-called BHL-prescription \cite{bhl,bhl2}, which is obtained by using the connection between the equal-time wavefunction $\psi_{c.m.}(\mathbf{q}_\perp)$ in the rest frame and the light-cone wavefunction $\psi_{LC}(x,\mathbf{k}_\perp)$ in the infinite momentum frame, i.e. $\psi_{c.m.}(\mathbf{q}_\perp) \leftrightarrow \psi_{LC}\left(\frac{\mathbf{k}^2_\perp+m^2} {4x(1-x)}-m^2\right)$, where $m$ stands for the light constitute quark mass. It is found that the the transverse momentum dependence is just in the exponential form of the off-shell energy of the constitute quarks, which agree with the Brodsky and Teramond's holographic model that is obtained by using the anti-de Sitter / conformal field theory correspondence \cite{wfbrodsky1,wfbrodsky2}. And, the pionic twsit-3 wavefunctions $\psi^\pi_{p}(x,\mathbf{k}_\perp)$ and $\psi^\pi_{\sigma}(x,\mathbf{k}_\perp)$ take the following form
\begin{widetext}
\begin{equation}
\psi^\pi_p(x,\mathbf{k}_\bot)=\left[1+B_p C^{1/2}_2(2x-1) +C_p C^{1/2}_4(2x-1) \right] \frac{A_p}{x(1-x)} \textrm{exp} \left[ -\frac{m^2 + \mathbf{k}^2_\bot} {8\beta_p^2x(1-x)} \right]
\end{equation}
and
\begin{equation}
\psi^\pi_\sigma(x,\mathbf{k}_\bot) = \left[ 1+B_\sigma C^{3/2}_2(2x-1) + C_\sigma C^{3/2}_4(2x-1) \right] \frac{A_\sigma}{x(1-x)} \textrm{exp} \left[ -\left(\frac{m^2+\mathbf{k}^2_\bot}{8\beta_\sigma^2 x(1-x)}\right)\right].
\end{equation}
\end{widetext}
The parameters $A_{p,\sigma}$, $B_{p,\sigma}$, $C_{p,\sigma}$ and $\beta_{p,\sigma}$ can be determined by the average value of the transverse momentum $\left< \mathbf{k}^2_\bot \right>_{p,\sigma}$, the wavefunction normalization $\int^1_0 dx \int_{|\mathbf{k}_\bot|<\mu_f}\frac{d^2\mathbf{k}_\bot} {16\pi^3} \psi^\pi_{p,\sigma}(x,\mathbf{k}_\bot)=1$, and the first two DA moments determined by the last section. The average value of the transverse momentum square is defined as
\begin{eqnarray}
\left< \mathbf{k}^2_\bot \right>_{p,\sigma} = \frac{\int dx d^2 \mathbf{k}_\bot \left| \mathbf{k}^2_\bot \right| \left| \psi^\pi_{p,\sigma} (x, \mathbf{k}_\bot) \right|^2}{\int dx d^2\mathbf{k}_\bot \left| \psi^\pi_{p,\sigma}(x,\mathbf{k}_\bot) \right|^2}
\end{eqnarray}
and $\left< \mathbf{k}^2_\bot \right>^{1/2}_{p,\sigma}=0.350\rm GeV$ \cite{PRD43}. Using the relation between the DA and the wavefunction, we obtain
\begin{widetext}
\begin{equation}
\phi^\pi_p(x,\mu_f) = \left[1+B_p C^{1/2}_2(2x-1) + C_p C^{1/2}_4(2x-1) \right] \frac{A_p \beta^2_p}{2\pi^2} \textrm{exp} \left[ -\frac{m^2}{8 \beta_p^2 x(1-x)} \right]\left\{ 1 - \textrm{exp} \left[ -\frac{\mu^2_f}{8 \beta_p^2 x(1-x)}\right] \right\}  \label{phiwp}
\end{equation}
and
\begin{equation}
\phi^\pi_\sigma(x,\mu_f) = \left[ 1+B_\sigma C^{3/2}_2(2x-1) + C_\sigma C^{3/2}_4(2x-1) \right] \frac{A_\sigma \beta^2_\sigma}{2\pi^2} \textrm{exp} \left[ -\frac{m^2}{8 \beta_\sigma^2 x(1-x)} \right] \left\{ 1 - \textrm{exp} \left[ -\frac{\mu^2_f}{8 \beta_\sigma^2 x(1-x)}\right] \right\}. \label{phiws}
\end{equation}
\end{widetext}

\begin{table*}
\centering
\caption{Wavefunction parameters for $\psi^\pi_p(x,\mathbf{k}_\perp)$ with varying DA moments. }
\begin{tabular}{|c||c|c|c|c|c|c|c|c|c|}
\hline
$\left< \xi^2_p \right>$  & \multicolumn{3}{|c|}{$0.196$} & \multicolumn{3}{|c|}{$0.248$} & \multicolumn{3}{|c|}{$0.324$}\\ \hline
$\left< \xi^4_p \right>$  & $0.207$ &  $0.262$ & $0.342$ & $0.207$ &  $0.262$ & $0.342$ & $0.207$ &  $0.262$ & $0.342$ \\ \hline\hline
$A_{p}(GeV^{-2})$ & 93.4987 & 92.4429  & 90.9983 & 89.7356 & 88.7233  & 87.2936 & 84.7201 & 83.7330  & 82.3809 \\ \hline
$B_{p}$           & 1.2835  & 1.2926   & 1.3056  & 1.3177  & 1.3261   & 1.3385  & 1.3679  & 1.3756   & 1.3868  \\ \hline
$C_{p}$           & 1.3579  & 1.4165   & 1.5005  & 1.3635  & 1.4210   & 1.5045  & 1.3710  & 1.4277   & 1.5097  \\ \hline
$\beta_{p} (GeV)$ & 0.5849  & 0.5882   & 0.5930  & 0.5950  & 0.5984   & 0.6033  & 0.6095  & 0.6131   & 0.6181  \\ \hline
\end{tabular}
\label{phip-par}
\end{table*}

\begin{table*}
\centering
\caption{Wavefunction parameters for $\psi^{\pi}_{\sigma}(x,\mathbf{k}_\perp)$ with varying DA moments. }
\begin{tabular}{|c||c|c|c|c|c|c|c|c|c|}
\hline
$\left< \xi^2_\sigma \right>$  & \multicolumn{3}{|c|}{$0.077$} & \multicolumn{3}{|c|}{$0.102$} & \multicolumn{3}{|c|}{$0.127$}\\ \hline
$\left< \xi^4_\sigma \right>$   & $0.074$ &  $0.094$ & $0.122$ & $0.074$ &  $0.094$ & $0.122$ & $0.074$ &  $0.094$ & $0.122$ \\ \hline\hline
$A_{\sigma}(GeV^{-2})$          & 196.486 & 194.804  & 192.128 & 187.336 & 185.532  & 182.777 & 178.403 & 176.532  & 173.584 \\ \hline
$B_{\sigma}$                    & -0.0284 & -0.0309  & -0.0349 & -0.0115 & -0.0144  & -0.0191 & 0.0053  & 0.0019   & -0.0032 \\ \hline
$C_{\sigma}$                    & 0.0922  & 0.1108   & 0.1364  & 0.0885  & 0.1069   & 0.1322  & 0.0845  & 0.1026   & 0.1276  \\ \hline
$\beta_{\sigma} (GeV)$          & 0.4107  & 0.4123   & 0.4147  & 0.4169  & 0.4187   & 0.4213  & 0.4234  & 0.4253   & 0.4283  \\ \hline
\end{tabular}
\label{phisi-par}
\end{table*}

The Gegenbauer moments $\left< \xi^{2n}_{p,\sigma} \right>$ for the DAs defined by Eqs.(\ref{phiwp},\ref{phiws}) at the scale $\mu_f$ can be defined as
\begin{eqnarray}
\left< \xi^{2n}_p \right>|_{\mu_f} = \frac{\int^1_0 dx \phi^\pi_p(x,\mu_f) C^{1/2}_{2n}(2x-1)}{\int^1_0 dx [C^{1/2}_{2n}(2x-1)]^2}
\end{eqnarray}
and
\begin{eqnarray}
\left< \xi^{2n}_\sigma \right>|_{\mu_f} = \frac{\int^1_0 dx \phi^\pi_\sigma(x,\mu_f) C^{3/2}_{2n}(2x-1)} {\int^1_0 dx 6x(1-x) [C^{3/2}_{2n}(2x-1)]^2}.
\end{eqnarray}

With the help of the above formulae, we can determine the wavefunction parameters for varying DA moments, which are presented in TAB.\ref{phip-par} and TAB.\ref{phisi-par}.

\begin{figure}
\begin{center}
\includegraphics[width=0.35\textwidth]{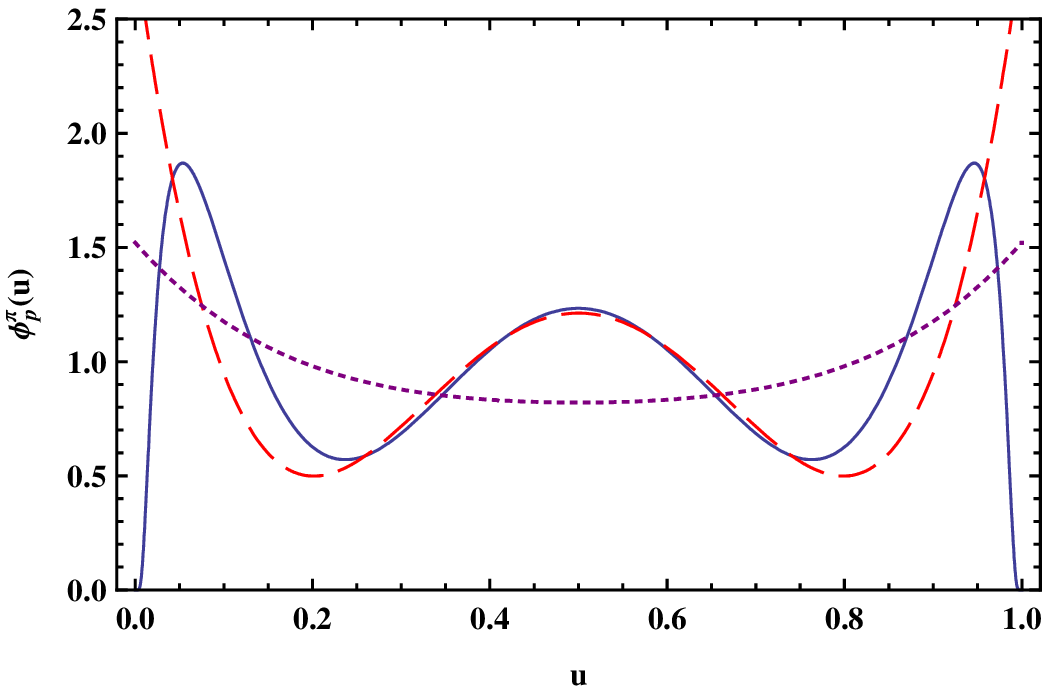}
\end{center}
\caption{Pion twist-3 DA $\phi^{\pi}_p$. The solid line, the dashed line and the dotted line are for our DA defined by Eq.(\ref{phiwp}), DA of Ref.\cite{ZPC48} and DA of Refs.\cite{NPB529,JHEP9901}, respectively. } \label{dis-p}
\end{figure}

\begin{figure}
\begin{center}
\includegraphics[width=0.35\textwidth]{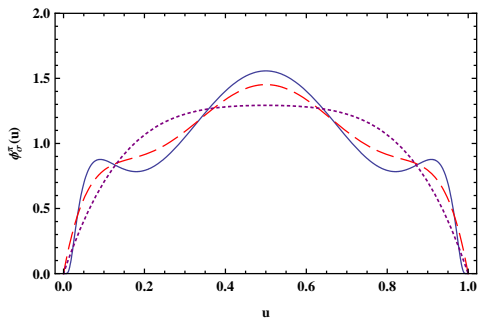}
\end{center}
\caption{Pion twist-3 DA $\phi^{\pi}_{\sigma}$. The solid line, the dashed line and the dotted line are for our DA defined by Eq.(\ref{phiws}), DA of Ref.\cite{ZPC48} and DA of Refs.\cite{NPB529,JHEP9901}, respectively. } \label{dis-sigma}
\end{figure}

We present $\phi^\pi_{p}$ and $\phi^\pi_{\sigma}$ in FIGs.(\ref{dis-p},\ref{dis-sigma}). Our DAs, i.e. Eqs.(\ref{phiwp},\ref{phiws}), are drawn by solid lines and are derived by setting the second and fourth moments to be their center values shown in TABs.(\ref{tabp1},\ref{tabp2},\ref{tabs1},\ref{tabs2}). As a comparison, we also present the DAs of Ref.\cite{ZPC48} and Refs.\cite{NPB529,JHEP9901} in FIGs.(\ref{dis-p},\ref{dis-sigma}), which are shown by the dashed lines and the dotted lines respectively. As for $\phi^\pi_p$, in different to its asymptotic behavior, it will provide great suppression in the end-point region. Such a behavior will be helpful to derive a reasonable power behavior for twist-3 DA \cite{bpiff}. It is noted that to study the right behavior of the DA itself, the full form of the twist-3 DAs (\ref{phiwp}) and (\ref{phiws}) is more useful and more accurate than its truncated Gegenbauer form.

\subsection{reanalysis of the $B\to\pi$ transition form factor}

The $B\to\pi$ transition form factor provides a good platform to study the pion properties. Especially, within the QCD light-cone sum rules \cite{sr}, one can concentrate on different twist structures of pion DA by properly choosing the correlator. Since by choosing the corrector with proper current, because of the cancelation among the different $\gamma$-structures, different twist DAs will be remained in the final formulae.

Three typical correlators are suggested in the literature to calculate $f^+_{B\pi}(q^2)$. The first one is $\Pi^{+}_\mu (p,q)=i\int d^4xe^{iq\cdot x} \langle \pi(p)|T \{ \overline{q}(x)\gamma _\mu b(x), \overline{b}(0) i m_b \gamma_5 q(0)\}|0\rangle$, which is usually adopted in the literature. After doing the simplification, it can be found that all the twist-2 and twist-3 and twist-4 terms are remained and all of which can provide sizable contributions \cite{alltwist,alltwist1,alltwist2}. The second one is $\Pi^{+}_\mu (p,q) =i\int d^4xe^{iq\cdot x}\langle \pi(p)|T \{\overline{q}(x)\gamma _\mu (1+\gamma _5)b(x), \overline{b}(0) i m_b(1+\gamma _5)q(0)\}|0\rangle$, which is constructed by using the chiral current. After doing the simplification, it can be found that the twist-3 terms can rightly be canceled, so one only needs to consider the twist-2 and twist-4 DAs \cite{chiral,chiral1}. The third one is $\Pi^{+}_\mu (p,q) =i\int d^4xe^{iq\cdot x}\langle \pi(p)|T \{\overline{q}(x)\gamma _\mu (1+\gamma _5)b(x), \overline{b}(0) i m_b(1-\gamma _5) q(0)\}|0\rangle$. After doing the simplification, it can be found that the terms involving the twist-2 DA are exactly be canceled, so one only needs to consider the dominant twist-3 DAs \cite{ZWH}.

In the present paper, our purpose is to test the properties of the obtained twist-3 DAs in the last section, so we adopt the third correlator to do our analysis. Following the same procedure of Refs.\cite{ZWH,alltwist2}, we can obtain the light-cone sum rules for the $f^+_{B\pi}(q^2)$ up to NLO,
\begin{widetext}
\begin{eqnarray}
f^+_{B\pi}(q^2) &=& \frac{f_\pi\mu_\pi}{m^2_B f_B}\exp\left[\frac{m^2_B}{M^2}\right] \left[ F_0 (q^2,M^2,s_0) + \frac{\alpha_s C_F}{4 \pi} F_1 (q^2,M^2,s_0) \right],
\label{twist3}
\end{eqnarray}
\end{widetext}
where the leading order $F_{0}(q^2,M^2,s_0)$ takes the form
\begin{widetext}
\begin{eqnarray}
F_0 (q^2,M^2,s_0) &=& m_b \left\{\int^1_{\Delta} \frac{du}{u} \exp\left[-\frac{m^2_b - q^2(1-u)}{uM^2}\right] \left[ u\phi_p^\pi(u) +  \frac{1}{6}\left(2+\frac{m^2_b+q^2}{uM^2}\right) \phi_{\sigma}^\pi(u)\right]-  \right. \nonumber\\
&&\!\!\!\!\!\!\!\!\!\!\!\!\!\!\!\!\!\!\!\!\!\!\!\! \left. \frac{2f_{3\pi}}{f_\pi\mu_\pi}\int^1_0 v dv \int D\alpha_i \frac{\theta(\alpha_1 +v\alpha_3 - \Delta)}{(\alpha_1 +v\alpha_3)^2} \exp\left[- \frac{m^2_b - q^2(1-\alpha_1-v\alpha_3)}{(\alpha_1+v\alpha_3)M^2}\right]
\left[1-\frac{m^2_b-q^2}{(\alpha_1+v\alpha_3)M^2}\right]\phi_{3\pi}(\alpha_i) \right\}
\end{eqnarray}
\end{widetext}
and the NLO $F_{1}(q^2,M^2,s_0)$ takes the form
\begin{widetext}
\begin{eqnarray}
F_1(q^2,M^2,s_0)=\frac{1}{\pi m_b} \int^{s_0}_{m^2_b} ds \  e^{-s/M^2} \int^1_0 du \left[ \textrm{Im}_s T^p_1 (q^2,s,u) \phi^\pi_p (u) + \textrm{Im}_s T^\sigma_1 (q^2,s,u) \phi^\pi_\sigma (u) \right] ,
\end{eqnarray}
\end{widetext}
where $D\alpha_i=d\alpha_1d\alpha_2 d\alpha_3 \delta(1-\alpha_1-\alpha_2-\alpha_3)$, $m_b$ stands for the running b-quark mass under the $\overline{MS}$ scheme, $\triangle=\frac{m^2_b-q^2} {s_0-q^2}$, $\phi_{3\pi}(\alpha_i)$ is the pionic three-particles twist-3 DA with the normalization parameter $f_{3\pi}$. The leading order result agree well with that of Ref.\cite{ZWH}. And it has been found that our present NLO result is different to that of Ref.\cite{alltwist2} only by an overall factor `2', which is caused by the use of different correlators. So to short the paper, we do not present them here, and the interesting readers may turn to Ref.\cite{alltwist2} for detailed formulae. Moreover, the three-particles twist-3 DA can be derived from the exact relation between the two-particle and three-particle twist-three DAs, which can be derived by the equations of motion \cite{ZPC48}. It is noted that the contributions from $\phi_{3\pi}(\alpha_i)$ to the present sum rules (\ref{twist3}) is quite small, which is less than $0.5\%$, so we simply take its form as suggested in Ref.\cite{ZPC48} to do our calculation.

\begin{figure}
\begin{center}
\includegraphics[width=0.5\textwidth]{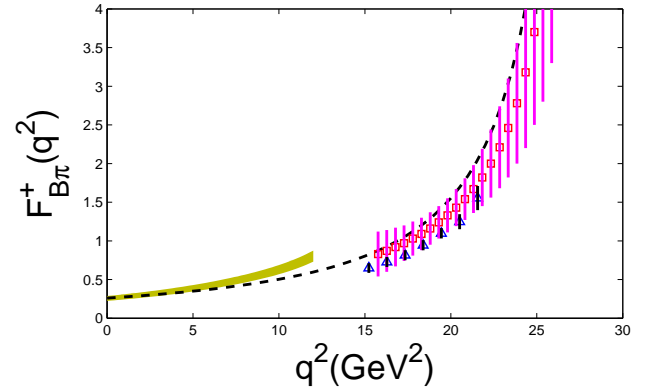}
\end{center}
\caption{ $f^+_{B\pi}(q^2)$. The shaded band is our result with uncertainties caused by the input parameters, and as a comparison, the extrapolated light-cone sum rules of Ref.\cite{chiral1} is shown by a dashed line. The quenched \cite{PRD64} and unquenched \cite{PRD73} lattice QCD results are presented by rectangles and triangles respectively. }
\label{formfactor}
\end{figure}

We present our results for $f^+_{B\pi}$ in FIG.(\ref{formfactor}), where the light-cone sum rule with certain extrapolation \cite{chiral1}, and the quenched lattice QCD result \cite{PRD64} and the unquenched lattice QCD result \cite{PRD73} are also presented for comparison. The light-cone sum rule for $B\to\pi$ is valid up to momentum transfer $q^2\sim m_b^2-2m_b\bar{\Lambda}$, typically at $0<q^2<14$ GeV$^2$, and to be on safe side, we take the maximal allowed $q^2$ to be $12$ GeV$^2$. Here in doing the numerical calculation, the center value of $f^+_{B\pi}$ is obtained by taking: $f_B=169$ MeV, $s_0=32.6$ GeV$^2$ \cite{chiral1} and $\bar{m}_b (\bar{m}_b)=4.20$ GeV, $m_B = 5.279$ GeV \cite{pdg}. The effective threshold for different correlators should be differ from each other so as to derive a reliable estimation \cite{s0}. As for the present adopted correlator with chiral current, the value of $s_0$ should be slightly lower than that of the conventional corrector (i.e. the first correlator) so as to include the unwanted scalar state with $J^P=0^+$ into the continuum contribution \cite{s02}.

It is found that our present result agrees with those of Refs.\cite{ZPC48,NPB529,JHEP9901} under the same parameter values. The shaded band is obtained by varying the pionic parameters within their reasonable regions. As for the Borel parameter, its reasonable region is $[14,20]$ GeV$^2$ and we fix its value to be $M^2=16$ GeV$^2$ to do our calculation. The typical energy scale for $B\to\pi$ is $\mu_b = (m^2_B - m^2_b)^{1/2} \simeq 2.2$ GeV, and by running $\mu_\pi$ from $1$ GeV to $\mu_b$, we obtain $\mu_\pi(\mu_b)=1.93\pm 0.24$ GeV. At $q^2=0$, we obtain $f^+_{B\pi}(0)=0.253\pm 0.031$. It is found that the main uncertainty is caused by $\mu_\pi$, and the twist-3 DAs models (\ref{phiwp}) and (\ref{phiws}) lead to small uncertainties. As for the NLO corrections, it will provide sizable contributions and its vaule will increase with the increment of $q^2$, e.g. for $q^2=0$ its contribution is $\sim 6\%$ and for $q^2=10$ GeV$^2$ its contribution changes to $\sim 16\%$, so it should be included to provide a sound estimation. As a cross check, it is found that our present result of $f^+_{B\pi}(0)$ is consistent with those of Refs.\cite{alltwist1,alltwist2,chiral,chiral1} within reasonable parameter regions.

\hspace{2cm}

\section{Summary}

We have studied the twist-3 DAs $\phi^{\pi}_{p}$ and $\phi^{\pi}_{\sigma}$ within the QCD sum rules. And by adding all the uncertainties in quadrature, the uncertainties for the moments are about $20\%-35\%$. Further more, these moments can provide helpful constraints on the twist-3 DAs and hence the wavefunctions. The present constructed twist-3 wavefunctions shall be useful for the $k_T$ factorization approach \cite{kt}, where the transverse momentum dependence both in the wavefunction and the hard scattering part should be treated on equal footing. A consistency check of our twist-3 DAs is done by studying the $B\to\pi$ transition form factor $f^+_{B\pi}$ with the QCD light-cone sum rules up to NLO.

Recently, Ref.\cite{btwist} points out a non-negligible contribution of higher-twist processes in large $p_t$ hadron production in hadronic collisions, where the hadron is produced directly in the hard subprocesses rather than by gluon or quark jet fragmentation. So a better understanding of higher twist wavefunctions or DAs shall be helpful for further studies on higher twist contributions. Especially, the forthcoming RHIC and LHC measurements will provide further tests of the dynamics of large-$p_t$ hadron production beyond the leading twist.

\hspace{2cm}

{\bf Acknowledgements}: This work was supported in part by the Fundamental Research Funds for the Central Universities under Grant NO.CDJXS10100035, by Natural Science Foundation Project of CQ CSTC under Grant No.2008BB0298 and by Natural Science Foundation of China under Grant No.10805082 and No.11075225.


\begin{thebibliography}{}

\bibitem{brodsky} G.P. Lepage and S.J. Brodsky, Phys.Rev. D{\bf 22}, 2157(1980); S.J. Brodsky and G.P. Lepage, Phys.Rev. D{\bf 24}, 1808(1981)

\bibitem{cz} V.L. Chernyak and A.R. Zhitnitsky, Nucl.Phys. B{\bf 201}, 492(1982).

\bibitem{babar} B. Aubert, et al., BABAR Collaboration, Phys.Rev. D{\bf 80}, 052002(2009).

\bibitem{twist2DA} X.G. Wu and T. Huang, Phys.Rev. D{\bf 82}, 034024(2010); A.V. Radyushkin, Phys.Rev. D{\bf 80}, 094009(2009); And references there in.

\bibitem{pballt} P. Ball, V.M. Braun and A. Lenz, JHEP{\bf 0605}, 004(2006).

\bibitem{NPB529} P. Ball, V. M. Braun, Y. Koike and K. Tanaka, Nucl.Phys. B{\bf 529}, 323(1998).

\bibitem{JHEP9901} P. Ball, JHEP{\bf 9901}, 010(1999).

\bibitem{hw1} A. Szczepaniak, A.G. Williams, Phys.Lett. B{\bf 302}, 87(1993); B.V. Geshkenbein, M.V. Terentyev, Phys.Lett. B{\bf 117}, 243(1982).

\bibitem{hw2} C.S. Huang, Commun.Theor.Phys. {\bf 2}, 1265(1983).

\bibitem{hw3} V.L. Chernyak, A.R. Zhitnitsky, Phys. Rep. {\bf 112}, 173(1984); B.V. Geshkenbein, M.V. Terentev, Sov.J.Nucl.Phys. {\bf 39}, 873(1984).

\bibitem{hw4} F.G. Cao, Y.B. Dai and C.S. Huang, Euro.Phys.J. C{\bf 11}, 501(1999).

\bibitem{hw5} Z.T. Wei and M.Z. Yang, Phys.Rev. D{\bf 67}, 094013(2003).

\bibitem{kt} H.N. Li and G. Sterman, Nucl.Phys. B{\bf 381}, 129(1992); J. Botts and G. Sterman, Nucl.Phys. B{\bf 325}, 62(1989).

\bibitem{PRD70} T. Huang, X. H. Wu and M. Z. Zhou, Phys.Rev. D{\bf 70}, 014013(2004).

\bibitem{EPJC42} T. Huang, M. Z. Zhou and X. H. Wu, Eur.Phys.J. C{\bf 42}, 271(2005).

\bibitem{piff} T. Huang and X.G. Wu, Phys.Rev. D{\bf 70}, 093013(2004); U. Raha and A. Aste, Phys.Rev. D{\bf 79}, 034015(2009).

\bibitem{bpiff} T. Huang and X.G. Wu, Phys.Rev. D{\bf 71}, 034018(2005); T. Huang, C.F. Qiao and X.G. Wu, Phys.Rev. D{\bf 73}, 074004(2006); T. Kurimoto, Phys.Rev. D{\bf 74}, 014027(2006).

\bibitem{YF41} A. R. Zhitnitsky, I. R. Zhitnitsky and V. L. Chernyak, Yad.Fiz. {\bf 41}, 445(1985).

\bibitem{ZPC48} V. M. Braun and I. E. Filyanov, Z.Phys. C{\bf 48}, 239(1990).

\bibitem{bkg1} V.A. Novikov, M.A. Shifman, A.I. Vainshtein and V.I. Zakharov, Fortschr.Phys. {\bf 32}, 585(19984).

\bibitem{bkg2} W. Hubschmid and S. Mallik, Nucl.Phys. B{\bf 207}, 29(1982).

\bibitem{bkg3} J. Govaerts, F.de Viron, D. Gusbin and J. Weyers, Phys.Lett. B{\bf128}, 262(1983).

\bibitem{bkg4} T. Huang, X.N. Wang and X.D. Xiang, Phys.Rev. D{\bf 35}, 1013(1987); T. Huang and Z. Huang, Phys.Rev. D{\bf 39}, 1213(1989).

\bibitem{condensates} P. Colangelo and A. Khodjamirian, arXiv:0010175.

\bibitem{sch} M. A. Shifman, Nucl.Phys. B{\bf 173}, 13(1980).

\bibitem{pdg} C. Amsler et al., Particle Data Group, Phys.Lett. B{\bf 667}, 1(2008).

\bibitem{alphas} S. Bethke, Eur.Phys.J. C{\bf 64}, 689(2009).

\bibitem{sudakov} H.N. Li and H.L. Yu, Phys.Rev.Lett. {\bf 74}, 4388(1995); Phys.Lett. B{\bf 353}, 301(1995); Phys.Rev. D{\bf 53}, 2480(1996).

\bibitem{bhl} S.J. Brodsky, T. Huang and G.P. Lepage, in {\it Particles and Fields-2}, Proceedings of the Banff Summer Institute, Banff, Alberta, 1981, edited by A.Z. Capri and A.N. Kamal (Plenum, New York, 1983), P143; G.P. Lepage, S.J. Brodskyk T.Huang, and P.B. Mackenize, {\it ibid.}, p83; T. Huang, {\it in Proceedings of XXth International Conference on High Energy Physics}, Madison, Wisconsin, 1980, edited by L.Durand and L.G. Pondrom, AIP Conf.Proc.No. 69(AIP, New York, 1981), p1000.

\bibitem{bhl2} T. Huang, B.Q. Ma and Q.X. Shen, Phys.Rev. D{\bf 49}, 1490(1994).

\bibitem{wfbrodsky1} S.J. Brodsky and Guy F.de Teramond, Phys.Rev. Lett.{\bf 96}, 201601(2006); Phys.Rev. D{\bf 77}, 056007(2008).

\bibitem{wfbrodsky2} S.J. Brodsky and Guy F.de Teramond, arXiv:0802.0514.

\bibitem{PRD43} X. H. Guo and T. Huang, Phys. Rev. D{\bf 43}, 2931(1991).

\bibitem{sr} V.L. Chernyak and I.R. Zhitnitsky, Nucl.Phys. B{\bf 345}, 137(1990); I.I. Balitsky, V.M. Braun and A.V. Kolesnichenko, Nucl.Phys. B{\bf 312}, 509(1989).

\bibitem{alltwist} V.M. Belyaev, A. Khodjamirian and R. R\"{u}ckl, Z.Phys. C{\bf 60}, 349(1993). V. M. Belyaev, V. M. Braun, A. Khodjamirian and R. R\"{u}ckl, Phys.Rev. D{\bf 51}, 6177(1995)'.

\bibitem{alltwist1}P. Ball, JHEP 9809, 005(1998); P. Ball and R. Zwicky, JHEP0110, 019(2001); P. Ball and R. Zwicky, Phys.Rev. D{\bf 71}, 014015(2005).

\bibitem{alltwist2} G. Duplancic, A. Khodjamirian, Th. Mannel and B. Melic, JHEP0804, 014(2008).

\bibitem{chiral} T. Huang, Z. H. Li and X. Y. Wu, Phys.Rev. D{\bf 63}, 094001(2001); Z.G. Wang, M.Z. Zhou and T. Huang, Phys.Rev. D{\bf 67}, 094006(2003); T. Huang, Z.H. Li, X.G. Wu and F. Zuo, Int.J.Mod.Phys. A{\bf 23}, 3237(2008); X.G. Wu and T. Huang, Phys.Rev. D{\bf 77}, 074001(2009).

\bibitem{chiral1} X.G. Wu and T. Huang, Phys.Rev. D{\bf 79}, 034013(2009).

\bibitem{ZWH} M.Z. Zhou, X.H. Wu and T. Huang, High Energy Physics and Nuceal Physics {\bf 28}, 927(2004).

\bibitem{PRD64} S. Aoki, {\it etal.}, JLQCD Collaboration, Phys.Rev. D{\bf 64}, 114505(2001).

\bibitem{PRD73} E. Gulez, {\it etal.}, HPQCD Collaboration, Phys.Rev. D{\bf 73},074502(2006).

\bibitem{s0} Wolfgang Lucha, Dmitri Melikhov, Hagop Sazdjian and Silvano Simula, Phys.Rev. D{\bf 80}, 114028(2009); Wolfgang Lucha, Dmitri Melikhov and Silvano Simula, Phys.Atom. Nucl.{\bf 73}, 1770(2010).

\bibitem{s02} Z.H. Li, F.Y. Liang, X.Y. Wu and T. Huang, Phys.Rev. D{\bf 64}, 057901(2001); Z.H. Li, T. Huang, J.Z. Sun, Phys.Rev. D{\bf 66}, 076005(2002); T.M. Aliev, I. Kanik and A. Ozpineci, Phys.Rev. D{\bf 67}, 094009(2003); X.G. Wu, Y. Yu, G. Chen and H.Y. Han, arXiv:1002.0483.

\bibitem{btwist} F. Arleo, S.J. Brodsky, D.S. Hwang and A.M. Sickles, arXiv:0911.4604; F. Arleo, S.J. Brodsky, D.S. Hwang and A.M. Sickles, arXiv:1006.4045.

\end{thebibliography}
\end{document}